# Information Theoretic Approach to Social Networks*

**Oded Kafri**

We propose an information theoretic model for sociological networks. The model is a microcanonical ensemble of states and particles. The states are the possible pairs of nodes (i.e. people, sites and alike) which exchange information. The particles are the energetic information bits. With analogy to bosons gas, we define for these networks' model: entropy, volume, pressure and temperature. We show that these definitions are consistent with Carnot efficiency (the second law) and ideal gas law. Therefore, if we have two large networks: hot and cold having temperatures $T_H$ and $T_C$ and we remove Q energetic bits from the hot network to the cold network we can save W profit bits. The profit will be calculated from $W < Q\ (1-T_H/T_C)$, namely, Carnot formula. In addition it is shown that when two of these networks are merged the entropy increases. This explains the tendency of economic and social networks to merge.

## Introduction

The word network, like information, is overly used. For example, in Shannon's theory there are confusions originated from the fact that some conceive stored data as information, while Shannon information theory deals with a file transmission from a sender to a receiver. Similarly, we conceive a network as a static graphical diagram of links connecting nodes while actually a network is characterized by a flow between nodes. For example, the electrical networks are conceived as a static net of electrical cables connected together; transportation networks as a static net of roads and irrigation networks as a static net of pipes, etc. However, the flow of electricity, traffic or water is the essence of the networks. Many scientific papers ware published about various aspects of networks from Erdős [1] to Barabashi [2, 3, 4]. Many techniques were applied in these researches, from graph theory of Erdős and Rényi to load distribution and statistics [5]. These diverse approaches apply to the many aspects of networks. However, here we discuss another kind of networks which we call "social networks" which are in fact communication networks. In these networks we overlook the physical wiring between the nodes and focus solely on the flow between them. An example to such nets is the data networks. Most of the people in the world are connected somehow by physical data networks. Eventually, everyone can communicate with almost everybody. However, the flow of the voice signals between the people is varying constantly in time and not distributed uniformly among them. These networks are similar to a two dimensional fluid.

## Social Network: definition

In this paper we adopt a dynamic approach to nets. In our network there are $N$ nodes which can communicate with all the other nodes with no wiring limitations. Moreover, each connection between node $i$ and node $j$ has a value $R$ which is a measure of the flow intensity between the two nodes. For example, in social networks $R_{i,j}$ may be the number of communication channels used between node $i$ and node $j$. In economical network $R$ may represent the value of a transaction between two nodes.

Shannon, [6] in his 1949th paper "a mathematical theory pf communication", describes in a quantitative way how Bob communicates with Alice via transmitting a "file" to her. The file is a sequence of bits where each bit can be either zero or one. When Alice receives the sequence of bits, she explores their value and extracts the content that Bob sent her. Shannon defined the entropy of the files as the logarithm of the possible different contents that the sequence may contain. Since $N$ bits file has $2^N$ possible different contents, the entropy of the file is $N \ln 2$. Engineers are using, for their convenience, base 2 logarithm and therefore the Shannon entropy in this base is identical to the length of the file, $N$ bits.

Basically, Shannon's theory deals with a one way communication between a sender (Bob) and a receiver (Alice) in which the sender send one or several bits to a receiver. Bits carry uncertainty which is expressed by the entropy. After reading and interpreting the file, the receiver can find its content

In this paper we describe a group of $N$ senders. Each of these senders can send and also receive information from the other $N-1$ members of the group. We call the communication group of $N$ senders/ receivers a network. We also call each one of the $N$ senders/receiver a node. In addition, we call a one way single communication channel connecting $N_i$ to $N_j$ a link. We designate $R_{i,j}$ , as the number of links through which a sender $i$ can send messages to a receiver $j$. Similarly, $R_{j,i}$ designates the number of links used from $j$ to $i$. We assume that there is a total number of $R$ links in the network and $R$ can be any integer. The network can be described by a matrix:

$$\begin{pmatrix}
R_{1,N} & & & & & & & R_{N-1,N} & 0 \\
 & & & & & & & 0 & R_{N.N-1} \\
R_{1,j} & & & & R_{i,j} & & 0 & & \\
 & & & & & 0 & & & \\
 & & & & 0 & & & & \\
 & & & 0 & & & & & \\
R_{1,3} & & 0 & & & & & & \\
R_{1,2} & 0 & & & & & & & \\
0 & R_{2,1} & R_{3,1} & & R_{i,1} & & & & R_{N,1}
\end{pmatrix}$$

*Figure 1- networks matrix.*

Where $R = \sum_{j=1}^{N} \sum_{i=1}^{N} R_{i,j}$ .

The summation on a column $i$ is the total links outgoing from node $i$ to all other $N-1$nodes. Similarly, the summation on a row $j$ is the total links entering node $j$ from all other $N-1$ nodes.

The network described above is different from our standard visualization of a net as a static diagram. $R_{i,j}$ can vary constantly like a two dimensional fluid matrix having two constraints;

A. $R_{i,i} = 0$

B. $R_{i,j} \leq R$.

Communication and economic networks have such a dynamic nature. In this aspect one may compare the network to a two dimensional fluid in which there are constant nodes and energetic links that are i.e. pulses (classical harmonic oscillator) or any other logical quantity such as money, etc. The number of links may represent the bandwidth of the communication channel or the amount of the money transferred.

We can imagine the networks as a two dimensional boson gas with $K = N^2 - N$ states and $R$ particles. Therefore, we can calculate for it entropy, temperature, volume and pressure.

**Large Networks Statistics**

The number of microstates $W$ of boson gas of $R$ particles in $K$ states is given by

$$W = \frac{(R+K-1)!}{(K-1)!R!} \tag{1}$$

Planck [7] used this equation assuming that $(K + R) \gg 1$, and designating the "occupation number" $n \equiv \frac{R}{K}$, to obtain his famous result for the entropy;

$$S(R, K) = \ln W = K[(n + 1)\ln(n + 1) - n \ln n] \tag{2}$$

We define large network as a network in which $R \gg K$, In this network it is possible to remove energetic links from it with a negligible change in its statistical properties. The thermodynamic analogue to the large network is an infinite thermal bath.

The entropy of the large net is given by $S(R, K) = \ln W$ [8]. When one link is added, the entropy is given by,

$$S(R + 1, K) = \ln \frac{(R+K)(R+K-1)!}{(R+1)R!(K-1)!} = \ln \frac{R+K}{R+1} + S(R, K) \tag{3}$$

In the case that $R$ is a large number than $R/(R + 1) \approx 1$ and,

$$S(R + 1, K) \approx S(R, K) + \ln \frac{n+1}{n} \tag{4}$$

**Carnot Efficiency**

Suppose we have two large networks $H$ and $L$ having occupation numbers $n_H$ and $n_L$ . We remove $Q$ links from the $L$ net and put them in the $H$ net. If $n_H > n_L$ than the entropy of the $Q, L$ net links is higher than $Q$, $H$ net links, and the total entropy will be decreased. Therefore, we must add $W$ links to the $H$ net, in order to avoid entropy decrease such that,

$Q \ln[(n_L + 1)/n_L] \leq (Q + W) \ln[(n_H + 1)/n_H]$ or,

$$W \leq Q\{1 - \ln \frac{n_H(n_L+1)}{n_L(n_H+1)}\} \qquad (5)$$

In the case that $n_{H,}$ and $n_L \gg 1$ then,

$$W \leq Q(1 - \frac{n_L}{n_H}) \qquad (6)$$

Equation 6 is Carnot inequality for networks.

**Large Networks Temperature**

The definition of temperature is related to the definition of entropy. In classical heat engine the Carnot efficiency is,

$$W \leq Q(1 - \frac{T_L}{T_H}) \qquad (7)$$

Where $W$ is the work, $Q$ is the heat (energy removed or added) and $T$ is the temperature. The occupation number $n$ is related in the classical limit of blackbody radiation (photons) $n \gg 1$ to the temperature via,

$$nh\nu = k_B T \qquad (8)$$

Here $h$ is the Planck constant, $\nu$ is the oscillator frequency and $k_B$ is the Boltzmann constant.

Therefore, if we substitute for a constant frequency, $\nu$, in equation 6 we obtain equation 7.

We can calculate the temperature directly from,

$$T = \frac{Q}{S}$$

In equation 4 we obtained the entropy increase by adding one link $Q = 1$, namely,

$\varDelta S = \ln\frac{1+n}{n}$ .

Therefore,

$$T = \left(\ln\frac{1+n}{n}\right)^{-1} \tag{9}$$

This result can also be obtained from Planck equation (2),

$$T = \frac{\partial R}{\partial S}$$

$$\frac{\partial S}{\partial R} = \frac{1}{K}\frac{\partial}{\partial n}K[(n+1)\ln(n+1) - n\ln n] = \ln\frac{1+n}{n} = \frac{1}{T} \tag{10}$$

We see that the two ways yield the same result.

In the $\lim_{n\to\infty}\ln(1+\frac{1}{n}) = \frac{1}{n}$ and,

$$T = n \tag{11}$$

This result is consistent with equation 6.

**Large networks Volume**

The volume of the large net is the number of its states $K$. This is the major difference between a gas and a large net. In the linear world, and therefore in our intuition, the volume is an extensive quantity. However, in nets the number of nodes is the extensive quantity. Since $K_i = N_i(N_i$ -1) and $N$ is extensive, therefore $K$ is not extensive, i.e. when we combine two nets 1 and 2, $N = N_1 + N_2$ and,

$$V \equiv K = (N_1 + N_2)(N_1 + N_2 - 1) = K_1 + K_2 + 2N_1N_2$$

Or for large nets,

$$V \approx V_1 + V_2 + 2(V_1V_2)^{1/2} = \left(\sqrt{V_1} + \sqrt{V_2}\right)^2 \tag{12}$$

Namely, the volume of the combined net is greater than the sum of their volumes. This is a counterintuitive result. When we combine two networks there is an expansion as a result of the increase of the number of states. Combining nets at constant number of links (adiabatic process) results in cooling and entropy increase. This is an explanation to a known phenomenon that networks tend to merge. It is well known that entropy increase in adiabatic process does not exist in ideal gas thermodynamics.

**Large Networks Pressure**

The gas law states that the pressure of the gas multiplied by its volume is a measure of the energy of the gas. In our case the particles are identical. Therefore, the energy of the net is the number of its links.

$$\mathrm{P}V = R = VT\ , \tag{13}$$

where P is the pressure and $R$ is the number of the particles. With analogy we write

$$\mathrm{P} = \mathrm{T} \approx n \tag{14}$$

The pressure of a net is a measure of the tendency of two nets having different pressures to be combined together to equate their pressure and temperature to equilibrium, and thus to maximize the total entropy. Due to the non extensivity of the volume, the combined pressure of two nets may be lower than the pressure of each one of them separately.

**Large Networks Entropy**

From equation 2 for large nets

$$S = V\left[n\ln\left(\frac{1+n}{n}\right) + \ln(n+1)\right] \text{ or,}$$

$$S{=}V\lim_{n\to\infty}\left[\ln\left(\frac{1+n}{n}\right)^{n} + \ln(n+1)\right] \approx V[1+\ln(n+1)]\,) \approx V\ln(1+n) \tag{15}$$

**Example**

For example, we take two large nets 1 and 2 with known pressure and volume. We combine them together. What will be the pressure and volume of the final net?

The solution for ideal gases is simple:

$$\mathrm{P}_1V_1 + \mathrm{P}_2V_2 = (R_1 + R_2)k_BT$$

Or the temperature of the combined gases is,

$$T = \frac{\mathrm{P}_1V_1 + \mathrm{P}_2V_2}{(R_1 + R_2)k_B}$$

And the pressure of the final gas is,

$$\mathrm{P} = \frac{\mathrm{P}_1V_1 + \mathrm{P}_2V_2}{V_1 + V_2}$$

For nets, the result is affected by the non-extensive nature of the nets volume. The temperature is the occupation number. Since $R$ is extensive, therefore,

$R = R_1 + R_2 = n_1 V_1 + n_2 V_2$ and

$$T \approx \mathrm{P} \approx n \approx \frac{P_1 V_1 + P_2 V_2}{\left(\sqrt{V_1} + \sqrt{V_2}\right)^2} \qquad (16)$$

**Numerical example**

Suppose we have two nets, each with 50 nodes; one has occupation number of 50 and the other of 100. The two nets are combined. What will be the value of the thermodynamic quantities in equilibrium of these two combined nets?

The net law is $\mathrm{P}V = R$

Where P is the pressure=temperature=occupation number, $V$ is the number of states, $R$ is the number of links.

For net 1: $V_1 = 50 \times 49 = 2{,}450$ $R_1 = 2450 \times 50 = 122{,}500$ $T_1 = \mathrm{P}_1 = 50$

Foe net 2: $V_2 = 50 \times 49 = 2{,}450$ $R_2 = 2450 \times 100 = 245{,}000$ $T_2 = \mathrm{P}_2 = 100$

In the combined net: $V = 100 \times 99 = 9{,}900$ , $R = 367{,}500$ $T = \mathrm{P} = 37$

The entropy of net 1 is $S_1 = 2450[1 + \ln 51] = 12{,}034$, the entropy of net 2 is $S_2 = 2450[1 + \ln 101]$ =13,732, and the entropy of the combined net is $S = 9900[1 + \ln 38] = 45{,}648$. The entropy increase is then 19,882.

This result demonstrates the major difference between a net and an ideal gas. When we combine nets, the temperature and the pressure drop drastically as a result of the entropy increase originated from the states generation in the combined net. This exhibits the tendency of nets to combine.

**Summary and Applications**

Is there any value to thermodynamic analysis of networks? This question was probably asked about information theory 70 years ago. It was possible to send files from Bob to Alice without information theory. Actually Samuel Morse did it 100 years before Shannon's time.

However, the quantitative work of Shannon enables to find limits on file's compression. Similarly, thermodynamic analysis of networks has already proved itself to be useful in showing that the distribution of links in the nodes in large networks is Zipfian [9]. If we define the wealth of a node as the number of links that is has, we see that combining two nets does not increase the wealth but reduces the temperature. Reducing the temperature enables higher free links (free energy), and therefore higher data transfer on the same infrastructure. Equilibrium thermodynamics proved to be an important tool in engineering, chemistry and physics. Applying these tools to sociological networks dynamics may prove to be of some use. For example, defining temperature to a net may help in our understanding of data flow. Zipf distribution may help in finding the stable inequality of links [10].

In reference [8] a similar calculation was made for the entropy increase when a node is added to a net. The result obtained is similar to that of equation 15. Namely, each node generates about $2\ln(1+n)$ entropy. This result quantifies the entropic benefit of joining the crowd (high linkage nets or hot nets). In this paper we found that the entropy generation caused by adding a link to a net is $\ln(1+\frac{1}{n}) \approx \frac{1}{T}$. It means that with contradistinction to a node, a link will favor joining a network with lower linkage (colder net), which represents the tendency of links (energy) to flow from hot to cold. One should note that the entropy generation by adding link to a net is with accordance to Benford's law [11].

The concept of non-extensive volume can also describe an accelerated expansion without energy.

**Acknowledgments**

I thank H. Kafri for reading the MS and for helpful remarks.

**References**

***** This paper contains similar content to the paper by the same author (O.K.) "Money, Information and Heat in Networks Dynamics" published in "*Mathematical Finance Letters*" Vol 2014 (2014), Article ID 4. This paper was rewritten for the Information theory community.